# Correlated non-perturbative electron dynamics with quantum trajectories

**Ivan P. Christov**

*Physics Department, Sofia University, 1164 Sofia, Bulgaria*
*ipc@phys.uni-sofia.bg*

**Abstract:** An approach to *ab initio* electron correlation effects in atoms that uses quantum trajectories is presented. A comparison with the exact quantum mechanical results for 1D Helium atom shows that the major features of the correlated ground state distribution and of the strong field ionization dynamics are reproduced with quantum trajectories. The intra-atomic resonant transitions are described accurately by a trajectory ensemble. The present approach reduces significantly the computational time and it can be used for both bound and ionizing electrons.

### References and links

## 1. Introduction

Owning to the recent developments in laser technology large range of intensities and pulse durations became available to the experiment. The field strengths produced in the laboratories can exceed significantly the Coulomb field experienced by the electrons in the atom, while the duration of the generated pulses span below one femtosecond. This opens broad avenues for investigation of the detailed electron motion in atoms and molecules that may reveal new fundamental physics and is also of major importance for the design of ultra-small scale electronic and photonic devices. Therefore, progress in the calculation of time dependent quantum-mechanical effects in atomic, molecular, and condensed matter systems has become indispensable. The reason is that the many-electron systems are in general described by a multi-partite time dependent wave function $\Psi(x_1,x_2,\ldots,x_n,t)$ whose practical computation is extremely demanding because the computational effort scales exponentially with the system dimensionality. Therefore, only atoms with a few electrons can be treated by a direct solution of time dependent Schrödinger equation. Several approximate methods have been used in order to make the time dependent evolution more tractable. One of those is the single active electron (SAE) approximation where only the most outer electron interacts with the laser field [1]. However, for strong fields and in attosecond time scale it is expected that the detailed motion of one of the electrons can significantly modify the motion of the rest of the electrons in their orbits, so clearly such correlation effects cannot be taken into account by SAE. Two other approaches that go beyond SAE are the time dependent Hetree-Fock method and the time dependent density functional method, but these also fail to describe correct correlated dynamics in strong fields [2-4]. More recently, much progress has been achieved by multi-configurational time-dependent Hartree [5] (-Fock [6]) method which allows the inclusion of correlation effects in a more systematic way. However, these methods also seem to be time-expensive when dealing with multi-electron systems. An alternative approach that accounts for the correlated electron motion in the regime of strong field ionization uses classical equations of motion instead of the Schrödinger equation. It has proven to be able to reproduce well the atomic non-sequential double ionization, with an order of magnitude less computational effort [7]. Clearly, the use of classical methods is limited when quantum interference effects are important. Here, a quantum trajectory method is used which combines the promise of delivering exact quantum-mechanical results with the affordable time scaling of the classical calculations.

## 2. Model

We introduce the quantum trajectory method to high-field and short pulse interactions by using a simplified one-dimensional Helium model atom, with soft core potentials [8]. Accordingly, the time evolution of the two-electron wave function in an external field is governed by the Schrödinger equation (in atomic units):

$$i\frac{\partial}{\partial t}\Psi(x_1,x_2,t)=\left[H_0+iA(t)\left(\frac{\partial}{\partial x_1}+\frac{\partial}{\partial x_2}\right)\right]\Psi(x_1,x_2,t)\,, \tag{1}$$

where:

$$H_0=-\frac{1}{2}\frac{\partial^2}{\partial x_1^2}-\frac{1}{2}\frac{\partial^2}{\partial x_2^2}-\frac{2}{\sqrt{a^2+x_1^2}}-\frac{2}{\sqrt{a^2+x_2^2}}+\frac{1}{\sqrt{b^2+(x_1-x_2)^2}} \tag{2}$$

Although Eq. (1) can be solved numerically without further approximations, we apply here the standard mean-field approximation which reduces that equation to a set of equations for the single-electron orbitals. Provided that the wave function is factorized as $\Psi(x_1,x_2,t)=\varphi_1(x_1,t).\varphi_2(x_2,t)$, from Eq. (1) and Eq. (2) the Hartree equations are obtained:

$$i\frac{\partial}{\partial t}\varphi_i(x_i,t)=\left[-\frac{1}{2}\frac{\partial^2}{\partial x_i^2}-\frac{2}{\sqrt{a^2+x_i^2}}+\int dx_j\frac{|\varphi_j(x_j,t)|^2}{\sqrt{b^2+(x_i-x_j)^2}}+iA(t)\frac{\partial}{\partial x_i}\right]\varphi_i(x_i,t) \tag{3}$$

where *i,j=1,2,* and *j≠i*.

Unlike in the standard approaches (e.g. the density functional approximation) where the correlation effects are taken into account by adding additional terms to the Hamiltonian (e.g. in [3]), here we transform the mean-field Hartree term itself, so that the correlated electron dynamics can be included into the equations of motion. To this end we note first that the space-averaged character of the Hartree term stems from the integral over the coordinate $x_j$ which is also an argument of the time dependent wave function, whereas $x_j$ itself is a time-independent variable. In order to relate $x_{1,2}$ directly to the electron motion it is assumed here that there is an ensemble of $N$ couples of classical particles where each one of the particles belongs to a different orbital. Then, we assume that the $k$-th particle from the $j$-th electron orbital has a well defined trajectory $x_j^k(t)$, i.e. $\left|\varphi_j(x_j,t)\right|^2 \sim \delta[x_j - x_j^k(t)]$, where δ is the Dirac delta-function. After substituting in Eq. (3) we obtain the Schrödinger equation for the wave function $\varphi_i^k(x_i,t)$ which is associated to the $k$-th particle from the $i$-th electron orbital:

$$i\frac{\partial}{\partial t}\varphi_i^k(x_i,t) = \left[-\frac{1}{2}\frac{\partial^2}{\partial x_i^2} - \frac{2}{\sqrt{a^2+x_i^2}} + \frac{1}{\sqrt{b^2+[x_i - x_j^k(t)]^2}} + iA(t)\frac{\partial}{\partial x_i}\right]\varphi_i^k(x_i,t) \qquad (4)$$

where $i$=1,2; $k$=1,2,…N

In this way the Hartree equation has been transformed to a set of single-particle Schrödinger equations coupled by the time dependent *e-e* Coulomb potential (the third term in Eq. (4)) which includes explicitly the momentary position of the one electron from each couple. Clearly a particle-wave dualism has been assumed in that each electron is described by both trajectory and wave function. In order to complete that set of equations we need equation of motion for the particles, that would allow us to determine the trajectories $x_j(t)$. One intuitively clear and straightforward way is to use the relation between the velocity of the particle at a given space-time point and the wave function solution of the Schrödinger equation as provided by the de Broglie-Bohm mechanics [9]:

$$\frac{dx_i^k}{dt} \equiv v_i^k(x_i,t) = \mathrm{Im}\left[\frac{1}{\Psi(x_i,x_j,t)}\frac{\partial \Psi(x_i,x_j,t)}{\partial x_i}\right]_{x_i = x_i^k(t);\, x_j = x_j^k(t)} - A(t) \qquad (5)$$

The velocity field in Eq. (5) can be calculated by using the symmetry properties of the wave function. In this way the exchange interaction can be accounted for without additional terms in the Schrödinger equation. For example, the field free Hamiltonian in Eq. (1) is symmetric under the exchange of the two coordinates $x_1$ and $x_2$ and thus the wave function can be chosen to be symmetric (singlet state). Then, we have in Eq. (5) $\Psi(x_1,x_2,t) = \varphi_1(x_1,t)\varphi_2(x_2,t) + \varphi_1(x_2,t)\varphi_2(x_1,t)$. It is interesting to note that Eqs. (4,5) can be derived directly from Eqs. (1,2) by using an approximate polar decomposition of the wave function $\Psi(x_1,x_2,t) \approx R_1(x_1,t)R_2(x_2,t)\exp[iS(x_1,x_2,t)]$ where $R_{1,2}$ are the real-valued amplitudes and $S$ is the real-valued phase function. Then, following the standard approach of the de Broglie-Bohm theory one arrives at the Hamilton-Jacobi set of equations, which are nonlinear and contain quantum potentials [9]. Fortunately, these can be reverted back to the set of linear Schrödinger equations, Eq. (4), which are by far more favourable for numerical calculations.

## 3. Results

It is clear from the above considerations that the calculation of the correlated multi-electron dynamics can be simplified significantly since the trajectories used are finite number and these are contained in the physical space unlike in the original Schrödinger equation where the wave function is an object defined in the configuration space. Before we apply the quantum trajectory method to the correlated two-electron system we first implement it for a 1D single electron atom (Hydrogen) with a soft core potential. For Hydrogen all terms related to the second electron are removed from Eqs. (1-4). Our goal is to demonstrate how the quantum trajectory ensemble follows the time evolution of the atomic wave function during a resonant transition from the ground state to the first excited state of the model atom. First, the ground state is calculated so that the trajectory distribution matches the ground state probability given by the solution of the Schrödinger equation, as described below for Helium. The ground state distribution of the wave function and that of the trajectory ensemble is shown in the left hand image in Fig. 1. Next, the atom is irradiated by electric pulse with an amplitude 0.01 a.u. and frequency 0.395 a.u. which matches the resonant frequency of the electron transition. The duration of the pulse is chosen so that the atom is completely inverted to the first excited state after that pulse.

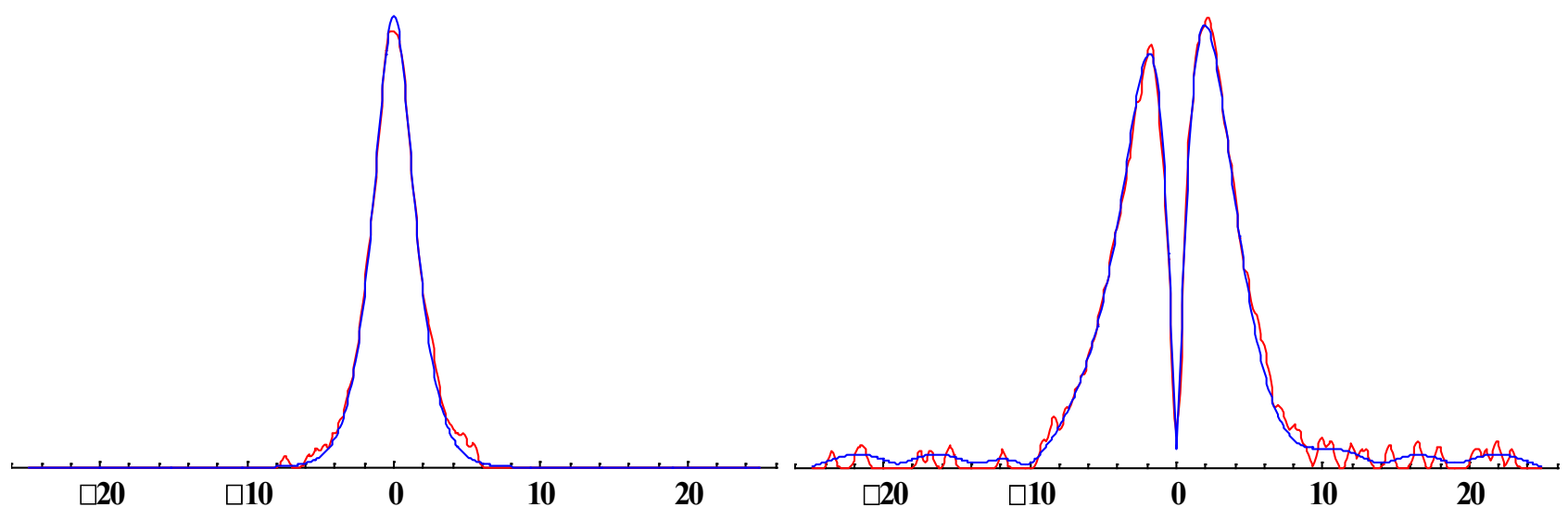

Fig. 1. Ground state of the Hydrogen 1D model atom (left). Blue-exact wave function, red-quantum trajectory distribution. The movie (531 KB) shows the evolution from the ground state to the first excited state (right). 10 000 quantum trajectories are used.

Next, we compare the results for the ground state of the 1-D Helium atom as calculated by the two-dimensional Schrödinger equation, Eq. (1) (for A(t)=0), and by using the trajectory approximation, Eqs. (4,5). The solution of the former is done by propagation of the initial Gaussian wave function in imaginary time until steady-state is established. The trajectory approach represents essentially a Monte-Carlo simulation where the initial ensemble consists of two randomly distributed sets of coordinates for the two electrons. These distributions are Gaussian and are chosen to match the probability distributions given by the initial wave functions $\varphi_1(x_1, t=0)$ and $\varphi_2(x_2, t=0)$. Then, for each couple of electrons from the ensemble the two Schrödinger equations (Eq. (4)) are propagated in complex time, while the trajectory equations (Eq. (5)) are solved in real time. Note that unlike in the standard approach complex time is used in Eq. (4) because if the wave function remains real-valued during the ground state calculation, the RHS in Eq. (5) would be zero at all times. As the steady state is approached the imaginary part of the ratio in the brackets in Eq. (5) converges to zero together with the velocities of the particles from the ensemble. As a result, the steady-state ground state wave-function matches the statistical distribution of the particle coordinates in the ensemble. Fig. 2 shows the position space distribution of 10 000 two-particle trajectories after steady state is reached, together with its interpolated version. It is seen that the approximate ground state distribution shows the characteristic “butterfly” shape of the contours where the dents along the $x_1=x_2$ line evidence the presence of an inner and an outer electron. This undoubtedly proves the correlated nature of our calculation for the ground state.

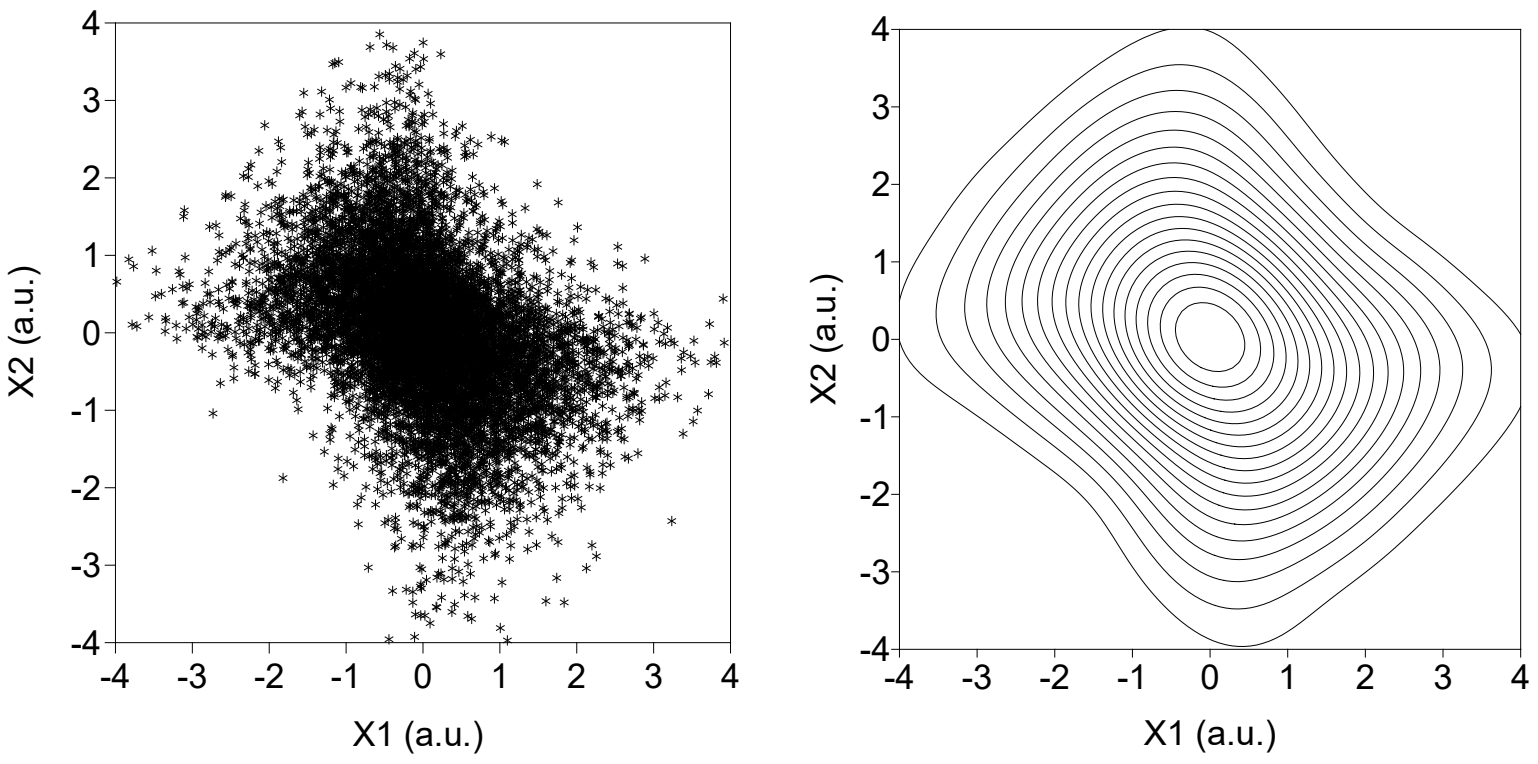

Fig. 2. Singlet ground state calculation for 1D Helium. The left hand image shows the distribution of 10 000 two-particle coordinates after steady-state of the quantum trajectories is established. The right hand image shows the same distribution after interpolation with Gaussians. Here a=b=1.

The dynamic behaviour of the quantum trajectory method is analyzed by irradiating the model atom by a six-cycle laser pulse at frequency ω=0.183 a.u. where the pulse is turned on and off linearly within two cycles and its central part is with constant amplitude. Fig.3 displays the results from the numerical solution of Eqs. (1,2), and from Eqs. (4,5), for the same parameters. As usual [2] the trajectories moving along the axes are interpreted as single ionization, while those far from both axes are associated with double ionization.

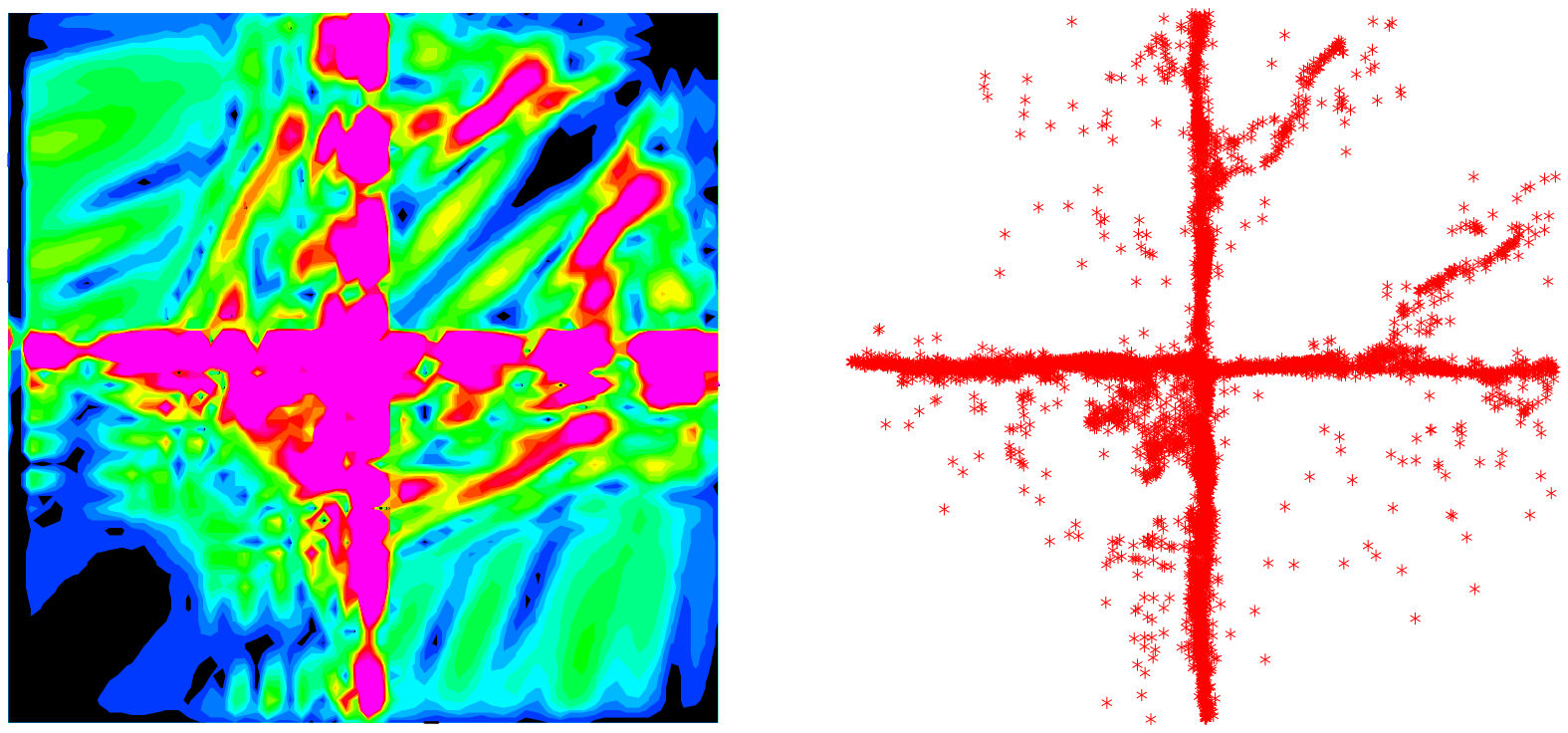

Fug. 3. Position space distributions for exact (left) and quantum trajectories (right) simulation of the strong field ionization of 1D Helium atom, three cycles after the pulse turn on. The movies associated with these pictures (1.3 MB, and 345 KB) show the time evolution of the two distributions. Here a=1, b=1.22.

The similarity between the exact result and the position space distribution from the trajectory simulations shown in Fig.3 is good, that is seen also from the attached movie. That similarity shows that the quantum trajectory calculation reproduces the essential correlated electron dynamics in the strong field regime.

An additional comparison between the quantum trajectories analysis and the exact model can be accomplished by calculating the double ionization probability as function of the laser intensity. The double ionization yield is defined as the probability of finding both electrons beyond some critical distance R from the core [4]. The results from the calculation of the double ionization probabilities are plotted in Fig.4 for R=12 a.u. It can be seen that the trajectory method gives results for the ionization yield that are close to the exact results over a range of four orders of magnitude. It was verified that for lower laser frequencies the quantum trajectories method gives somewhat enhanced ionization yield compared to the exact solution. This can be attributed to the one-dimensional model that we use where the particles within the correlated couples may get too close to each other in space, which causes instabilities in their motion and enhances the ionization. This effect is similar to 1D auto-ionization.

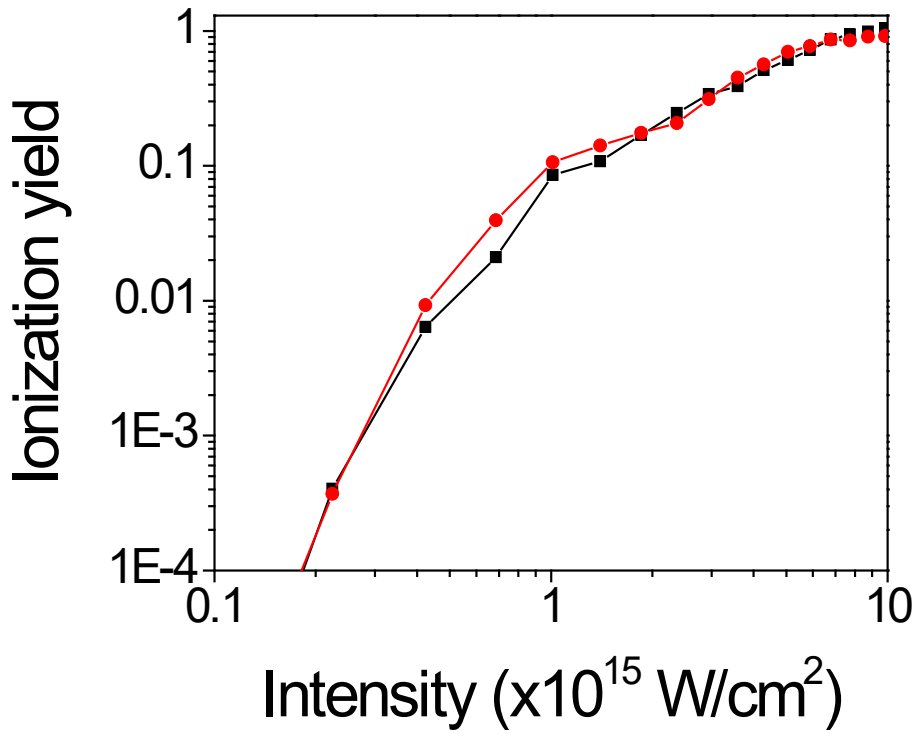

Fig. 4. The double-ionization yield calculated for six-cycle laser pulse at frequency 0.183 a.u. The red line shows the quantum trajectory result while the black line shows the exact result.

The method reported here can be applied to quantum systems with arbitrary number of electrons. The generalization to Hamiltonians that include spin and to relativistic calculations is straightforward. One major advantage of the quantum trajectories is that they provide visualization of the quantum phenomena which can help to better understand them. It should be noted that here we use the quantum trajectories as a convenient computational tool and ignore the interpretation issues connected with the Bohmian mechanics.

**Acknowledgment**

The author gratefully acknowledges support from the National Science Fund of Bulgaria under contract WUF-02-05.